\documentclass[11pt,a4paper]{article}

\usepackage{jcappub}
\usepackage{hyperref}
\usepackage{subfig}
\hypersetup{colorlinks = true, citecolor = blue}
\usepackage{cancel}

\begin{document}
%
%
\title{
Naturalness in Higgs inflation in a frame independent formalism
}
\author{Tomislav Prokopec}
\emailAdd{t.prokopec@uu.nl}
\author{and Jan Weenink}
\emailAdd{j.g.weenink@uu.nl}

\affiliation{Institute for Theoretical Physics, Spinoza Institute and\\
the Center for Extreme Matter and Emergent Phenomena (EMME$\Phi$),\\
Utrecht University, Leuvenlaan 4, 3585 CE Utrecht, The Netherlands}
\abstract{
 We make use of the frame and gauge independent formalism
for scalar and tensor cosmological perturbations developed in Ref.~\cite{Prokopec:2013zya}
to show that the physical cutoff for 2-to-2 tree level scatterings in Higgs inflation
is above the Planck scale $M_{\rm P}=1/\sqrt{8\pi G_N}$ throughout inflation.
More precisely, we found that in the Jordan frame, the physical cutoff scale
is $(\Lambda/a)_J\gtrsim \sqrt{M_{\rm P}^2 +\xi\phi^2}$, while in
the Einstein frame it is $(\Lambda/a)_J\gtrsim M_{\rm P}$, where $\xi$ is the nonminimal coupling
and $\phi$ denotes the Higgs {\it vev} during inflation. The dimensionless ratio of the physical cutoff
to the relevant Planck scale is equal to one in both frames, thus demonstrating
the physical equivalence of the two frames.
Our analysis implies that Higgs inflation is unitary up to the Planck scale, and
hence there is no naturalness problem in Higgs inflation.
In this paper we only consider the graviton and scalar interactions.
}
\keywords{}
\arxivnumber{}
\maketitle

\section{Introduction}
Higgs inflation \cite{Bezrukov:2007ep,Salopek:1988qh}
\cite{Bezrukov:2008ej,Barvinsky:2008ia,DeSimone:2008ei,Bezrukov:2009db,Barvinsky:2009fy,Barvinsky:2009ii,Bezrukov:2013fka} is perhaps the most
economical approach to cosmological inflation: it identifies the only
known scalar field in the Standard Model (SM) -- the Higgs boson -- with
the inflaton, the scalar field that drives the inflationary expansion
in the very early universe. The action for Higgs inflation
features a non-minimal coupling of the Higgs field to
the Ricci scalar. If the non-minimal coupling $\xi$ is large, it leads to
a successful period of chaotic inflation, producing primordial
power spectra that fit the observational bounds~\cite{Ade:2013uln}.
Hence, Higgs inflation can be considered as a "natural" scenario,
since it does not seem to require the introduction of new
physics to explain the inflationary expansion of the universe.

However, the naturalness of the Higgs inflation scenario has been
under debate. Refs.~\cite{Barbon:2009ya,Burgess:2009ea} used
power-counting techniques to determine the energy scale $\Lambda$
at which perturbation theory breaks down, thus determining the range
of validity of Higgs inflation. It was claimed that this cutoff scale
$\Lambda$ lies dangerously close to the energy scale of inflation,
thereby questioning the naturalness of Higgs inflation. Since then
many works have appeared claiming that Higgs inflation is
"natural"~\cite{Lerner:2009na,Ferrara:2010in}
or "unnatural"~\cite{Burgess:2010zq,Hertzberg:2010dc}.
Perhaps the most complete treatment has been done in Ref.~\cite{Bezrukov:2010jz}.
It was found that $\Lambda$ is generally field dependent and lies above the
typical energy scales in different regions, such that the perturbative
(semiclassical) expansion is valid in Higgs inflation.

Although there seems to be a consensus about the cutoff scale
as computed in Ref.~\cite{Bezrukov:2010jz} (however,
see the recent work~\cite{George:2013iia}~\footnote{
The results presented in Ref.~\cite{George:2013iia}
use different techniques and arrive at results that are consistent with
those presented in this paper and earlier in Ref.~\cite{Weenink:2013oqa}.}),
we revisit the computation of the cutoff in this work.
The reason is that there are some important aspects
that have not been fully taken into account. The most important
aspect is that General Relativity contains a large diffeomorphism symmetry,
which when truncated resembles the symmetry of a gauge theory.
This means that, like in QED, some of the degrees of freedom in the action are
actually not physical. As a consequence, some of
the interaction vertices obtained after a na\"\i ve perturbative
expansion of the action are gauge dependent, and any conclusion that one arrives at
by using these vertices can be a gauge artifact.
Moreover, it is possible that there are
additional vertices that conspire to cancel dominant
perturbative contributions. Accounting for these aspects can raise the cutoff scale.
Even though it is possible to determine the physical cutoff scale
within a gauge dependent formulation,
by far the simplest and most reliable way
to determine this scale is to use the physical vertices,
which can be obtained from the perturbative
action in a manifestly gauge invariant way (an alternative is to completely fix the gauge freedom
and take account of contributions from {\it all} vertices).
This formulation of the action in terms of \textit{gauge invariant perturbations}
has been found for a non-minimally coupled scalar field
up to third order in perturbations
in Refs.~\cite{Weenink:2010rr,Prokopec:2012ug,Prokopec:2013zya}.
In this work we use these previously found results
in order to demonstrate that the cutoff scale for physical, that is, gauge
invariant perturbations is always $\geq M_P$.
To be more precise, we find that
\begin{eqnarray}
\left(\frac{\Lambda}{a}\right)_J \gtrsim \sqrt{M_P^2+\xi\phi^2}
\,,\qquad
\left(\frac{\Lambda}{a}\right)_E  \gtrsim M_P
\,,
\label{cutoffscalesJordanEinstein}
\end{eqnarray}
where $a_J$ and $\phi$ are the background scale factor and scalar field
in the Jordan frame, and $a_E$ is the Einstein frame scale factor.
The extra scale factors have been overlooked in previous computations
of the cutoff scale. They appear since $\Lambda^{-1}$ is a comoving scale,
and therefore in an expanding universe the corresponding physical length scale,
$a/\Lambda$, always includes a scale factor $a$.
The cutoffs in Jordan and Einstein frame are different, simply
because $\Lambda$ is a dimensionful quantity which differs between
the frames, just like the effective Planck mass.
If $M_P$ is the energy scale where quantum gravity kicks in
in the Einstein frame, then $M_{P,J}\equiv \sqrt{M_P^2+\xi\phi^2}$ can
be identified with the scale of quantum gravity in the Jordan frame.
Thus in either frame the perturbative (semiclassical) treatment is
valid all the way up to the scale at which gravity becomes strong.
This means that Higgs inflation is perfectly natural and that
no new physics is necessary to explain the inflationary expansion
of the universe and the anisotropies in the CMB.

In order to arrive at the physical cutoff scales~\eqref{cutoffscalesJordanEinstein},
we first briefly discuss Higgs inflation and physically equivalent
frames in section~\ref{sec:Higgs inflation}. In section~\ref{sec: the naturalness debate}
we review the naturalness debate. In section~\ref{sec: gauge invariance}
we discuss the concept of gauge and frame dependence in cosmology
and quote from Ref.~\cite{Prokopec:2013zya}
the action for gauge and frame invariant cosmological perturbations.
Finally we compute the cutoff scale for physical
perturbations in section~\ref{sec: cutoffscale} and conclude in section~\ref{sec:Discussion}.

\section{Higgs inflation}
\label{sec:Higgs inflation}

Unfortunately, inflation in the pure SM
has been ruled out by observations of the CMB, which have put tight constraints on
the inflationary potential. However, these constraints
can be greatly relaxed when the Higgs field $\mathcal{H}$ is quadratically coupled
to the Ricci scalar $R$ by a large non-minimal coupling $\xi \sim 10^4$
\cite{Futamase:1987ua,Fakir:1990eg,Komatsu:1999mt,Tsujikawa:2004my}.
The pure scalar and gravitational part of the action reads
\begin{equation}
S = \int d^4x \sqrt{-g}
\biggl\{
\frac12 M_P^2 R +\xi\mathcal{H}^{\dagger}\mathcal{H} R
 - g^{\mu\nu} (\partial_{\mu}\mathcal{H})^{\dagger}(\partial_{\nu}\mathcal{H})
 -\lambda \left(\mathcal{H}^{\dagger}\mathcal{H}-\frac{v^2}{2}\right)^{\!2}
\biggr\}
\,.
\label{SMHiggsaction}
\end{equation}
Here the metric signature is ${\rm sign}[g_{\mu\nu}]=(-,+,+,+)$, the Higgs self-coupling is
$\lambda$ and the Higgs vacuum expectation value (vev) is $v=246~\rm{GeV}$.
Although the complex doublet $\mathcal{H}$ contains four degrees of freedom,
three of those can be absorbed by the gauge bosons (not shown in Eq.~\eqref{SMHiggsaction}).
In unitary gauge the remaining scalar degree of freedom with non-zero vev
can be parametrized by $\mathcal{H}=(0,\Phi/\sqrt{2})$,
such that the scalar action takes the Jordan frame form
\begin{equation}
S = \frac12 \int d^4x \sqrt{-g}
\Bigl\{
R F(\Phi) -g^{\mu\nu} (\partial_{\mu}\Phi)(\partial_{\nu}\Phi)
-\frac{\lambda}{2}(\Phi^2-v^2)^2
\Bigr\}
\,,
\label{Jordanframeaction}
\end{equation}
with $F(\Phi)=M_P^2+\xi\Phi^2$. The claim is now that
a successful period of (chaotic) inflation is possible
when the non-minimal coupling parameter is large, $\xi \sim 10^4$.
Successful means that the model's predictions for the primordial
power spectra of scalar and tensor perturbations
agree with the observational constraints on these spectra.
The current 1$\sigma$ constraints on the scalar spectral index
and the tensor-to-scalar ratio are $n_s = 0.9603\pm 0.0073$
and $r < 0.11$, respectively~\cite{Ade:2013uln}. On the theory
side, it is known that in a minimally coupled model ($\xi=0$)
both $n_s-1$ and $r$ are proportional to slow-roll parameters.
Moreover, since the Hubble parameter $H$ does not directly couple
to the time dependent expectation value of the scalar field
$\langle\Phi\rangle =\phi(t)$ in the Friedmann equations,
it is possible to express the slow-roll parameters in terms
of the slope of the potential. Hence, in a minimally coupled
model the shape of the scalar potential provides an intuition
for the successful realization of inflation. For instance,
a quartic potential is excluded by CMB observations~\cite{Ade:2013uln}
because it is insufficiently flat in the inflationary regime,
which supports the previously mentioned statement that inflation is
not possible in the pure (\textit{i.e.} minimally coupled) SM.

Unfortunately, we do not have the luxury of such an intuition
in the case of (large) non-minimal coupling $\xi \gg 1$.
The mixing of the gravitational and matter part of the action
prevents us from expressing the slow-roll parameters in terms
of the potential. Worse, it is not clear what are the slow-roll
parameters in a non-minimally coupled model,
that is, what are the parameters that remain small
during the inflationary period, and neither is it clear how
the primordial power spectra depend on these small parameters.
The most straightforward way to see whether or not Higgs inflation
works is therefore to compute the primordial power spectra
for non-minimal coupling from scratch, which involves a derivation of the quadratic
action for scalar and tensor perturbations in the Jordan frame~\cite{Weenink:2010rr}.
Fortunately we can avoid this exercise
by making use of a clever trick. If we redefine the metric,
scalar field and scalar potential in the action as follows~\cite{Weenink:2010rr},
\begin{align}
g_{\mu\nu,E}&=\frac{F}{M_P^2} g_{\mu\nu}
\nonumber\\
\left(\frac{d\Phi_E}{d\Phi}\right)^2
&=\frac{M_P^2}{F}\left(1+\frac{3}{2}\frac{F^{\prime 2}}{F}\right)
\nonumber\\
V_E(\Phi_E)&=\frac{M_P^4}{F^2}V(\Phi)
\,,
\label{Conformaltransformationmetricscalar}
\end{align}
where $F=F(\Phi)$, $F'=dF/d\Phi$ and $V(\Phi)=\frac14 \lambda (\Phi^2-v^2)^2$,
we find that the action \eqref{Jordanframeaction} becomes
\begin{equation}
S_E = \frac12 \int d^4x \sqrt{-g_E}
\Bigl\{
M_P^2 R_E -g_E^{\mu\nu} \partial_{\mu}\Phi_E\partial_{\nu}\Phi_E-2V_E(\Phi_E)
\Bigr\}
\,.
\label{Einsteinframeaction}
\end{equation}
Since the metric has been rescaled by a factor it is commonly said
that we are in another conformal frame; the specific frame for which
the scalar field is minimally coupled to the Ricci scalar
is called the Einstein frame, and quantities in this frame are indicated by a subscript $E$.
Thus the action has been rewritten
to the more familiar minimally coupled form. The crucial point
is that the Jordan and Einstein frame are related by
field redefinitions \eqref{Conformaltransformationmetricscalar}, and hence the
two formulations are physically equivalent. Nature is indifferent
to whether we use one set of variables to describe her phenomena,
or a different, but related, set. Physical observables should
therefore be invariant with respect to the frame in which you compute
them~\cite{Flanagan:2004bz}. Specifically, the primordial power spectra for scalar and tensor
perturbations (or the predicted values of $n_s$ and $r$) are the
same whether you compute them in the Jordan frame using variables
$g_{\mu\nu}$ and $\Phi$, or in the Einstein frame using $g_{\mu\nu,E}$
and $\Phi_E$ related to the first set by Eqs.~\eqref{Conformaltransformationmetricscalar}.
This was first demonstrated in Refs.~\cite{Makino:1991sg,Fakir:1992cg},
see also Ref.~\cite{Weenink:2010rr}. Hence, the more familiar and more intuitive
Einstein frame formulation can be used to check the successfulness of
Higgs inflation \cite{Bezrukov:2007ep,Salopek:1988qh}.
In the Einstein frame it is intuitively clear that Higgs inflation
works: the potential becomes exponentially flat in the large field limit.
Predictions for the spectral index and scalar-to-tensor ratio in Higgs inflation
are $n_s\simeq 0.97$ and $r \simeq 0.0032$, and are
therefore well within the observational bounds.

\section{The naturalness debate}
\label{sec: the naturalness debate}

At first sight Higgs inflation is a very attractive scenario, because
it does not seem to require the introduction of new physics
to explain the exponential expansion of the early universe.
However, this has been questioned by Refs.~\cite{Burgess:2009ea,Barbon:2009ya},
who looked at the ultraviolet cutoff scale $\Lambda$ in Higgs inflation.
$\Lambda$ indicates the energy scale above which scattering amplitudes
violate the unitarity bound, and therefore determines when perturbation
theory breaks down. Such a cutoff can be found by power-counting techniques
for which one can use the following recipe:
\begin{enumerate}
\item Perform a perturbative expansion of the fields in the action.
\item Rescale the fields such that their kinetic terms are canonically normalized.
\item Read off the cutoff scale from the interaction terms with $D>4$ which
are suppressed by $\Lambda^{4-D}$.
\end{enumerate}
In Refs.~\cite{Burgess:2009ea,Barbon:2009ya} the cutoff scale was obtained
by expanding the scalar field around its vev, $\Phi= v+\varphi$,
and the metric around Minkowski space, $g_{\mu\nu}=\eta_{\mu\nu} + h_{\mu\nu}/M_P$,
where the normalization $M_P^{-1}$ is chosen such that the gravitational
terms are canonically normalized.
In the Jordan frame, the cutoff follows from the expansion of the
$\xi \mathcal{H}^{\dagger} \mathcal{H} R$ term, which gives the
5-dimensional interaction $(\xi/M_P) \varphi^2 \Box h$. Reading off
the cutoff scale gives $\Lambda = M_P/\xi$. In the Einstein frame
the cutoff follows from a small-field expansion of the potential,
which gives $V_E=\frac14\lambda \varphi_E^4-\lambda(\xi/M_P)^2\varphi_E^6+\ldots$.
The dimension 6 terms are due to the small-field relation between Jordan and Einstein
frame fields $\Phi \simeq \Phi_E[1 - (\xi/M_P)^2]$.
Again, the cutoff scale from the dimension 6 term is $\Lambda=M_P/\xi$.
The breakdown of perturbation theory may signal the appearance of new
physics, for example higher dimensional operators suppressed by $\Lambda$,
which solve the unitarity problems at energy scales above $\Lambda$.
Now, the point is that the cutoff scale $\Lambda$ lies
dangerously close to the energy scale of inflation, characterized
by the Einstein frame Hubble parameter $H_E \simeq \sqrt{\lambda}M_P/\xi$.
These higher dimensional operators may therefore enter the inflationary
potential and affect inflationary predictions, or worse, spoil
Higgs inflation. Thus we need knowledge of the ultraviolet completion
of Higgs inflation in order to find out if inflation is successful,
which obviously clashes with the original attractiveness of the scenario.

Since the appearance of the works~\cite{Burgess:2009ea,Barbon:2009ya}
there has been a debate in literature about
the presence or absence of unitarity problems in Higgs inflation.
Refs.~\cite{Lerner:2009na,Ferrara:2010yw,Ferrara:2010in}
considered the Einstein frame analysis
and objected, correctly, that the dimension 6 terms
in the \textit{small field} expansion of the potential
only give rise to unitarity problems in the \textit{large field} regime
($\langle\Phi\rangle\gg M_P/\xi$ or $\langle \Phi_E \rangle \geq M_P$), where the small field
expansion is no longer valid. Instead, one should perform a
perturbative expansion around the field expectation value,
which results in a field dependent cutoff scale.

Then, by making use of the equivalence
between the Jordan and Einstein frame, it was argued that there are also
no unitarity problems in the Jordan frame for single field inflation.
Refs.~\cite{Burgess:2010zq,Hertzberg:2010dc} showed
subsequently that the unitarity bound $M_P/\xi$ again appears when
the Goldstone bosons are taken into account, both in the Jordan and Einstein
frame. Ref.~\cite{Burgess:2010zq} added that even in unitary gauge,
where the Goldstone bosons are eaten by the gauge bosons, the cutoff
scale appears in Higgs-gauge interactions. Ref.~\cite{Ferrara:2010yw,Ferrara:2010in}
argued that, instead of expanding around a small Higgs vev, the perturbative
expansion should be performed around a large expectation value $\phi\gg M_P/\xi$,
which is the background relevant for inflation. Here $\phi=\langle \Phi \rangle$,
with $\Phi$ the Higgs field in unitary gauge, $\mathcal{H}=(0,\Phi)$.
It was shown that the cutoff scale following from the Einstein frame
potential is $M_P$ in the inflationary regime,
but undergoes a transition to $M_P/\xi$ for small field values $\phi \ll M_P/\xi$. The authors
then argued that this post-inflation unitary bound does not affect our ability
to describe physical processes during inflation. Arguably the most
complete treatment of the cutoff so far has been performed
in Ref.~\cite{Bezrukov:2010jz}. There the metric was expanded around
an expanding background $g_{\mu\nu} =\bar{g}_{\mu\nu}+\delta g_{\mu\nu}$
and likewise the scalar field was expanded around a time dependent
background $\Phi=\phi+\varphi$. Next, the Jordan frame action was
put into canonical form by a redefinition of the metric and
scalar field perturbations.
In the Jordan frame the cutoff originates from the $\xi \Phi^2 R$ term
and was found to be $M_P/\xi$ for $\phi \ll M_P/\xi$,
$\xi\phi^2/M_P$ for $M_P/\xi < \phi < M_P/\sqrt{\xi}$ and $\sqrt{\xi}\phi$
for $\phi > M_P/\sqrt{\xi}$. In the Einstein frame the same results
were obtained from the potential \footnote{The cutoffs in the Jordan and Einstein
frames are related by the conformal factor $\sqrt{1+\xi\phi^2/M_P^2}$. This
can be understood by realizing that the cutoff is some energy scale, or length
scale, which is changed by the conformal transformation. In Ref. \cite{Bezrukov:2010jz}
the cutoffs are only found to be equivalent
(note: up to factors of the self-coupling $\lambda$)
once this factor is taken into
account, see also \cite{Bezrukov:2008ej,Bezrukov:2009db}.}.

\section{Gauge invariance}
\label{sec: gauge invariance}

In this work we revisit the computation of the cutoff scale
in Higgs inflation. The reason is that all of the papers
mentioned in the previous section have overlooked
a crucial aspect of the computation of quantum corrections in general relativity.
This crucial aspect is the fact that general relativity contains
a large diffeomorphism symmetry. Since the diffeomorphism symmetry
resembles many aspects of gauge symmetries, general relativity
is often called a gauge theory and its degrees of freedom (dofs)
are generally gauge dependent. Due to gauge dependence not all
dofs are physical. Moreover, general relativity is a constrained
theory, such that some of the dofs do not participate
in the dynamics, but instead impose constraints on the system.
Therefore, computing a cutoff by means of
expanding $g_{\mu\nu}=\bar{g}_{mu\nu}+\delta g_{\mu\nu}$
and $\Phi = \phi +\varphi$ and using the corresponding vertices,
will generally give non-physical and incorrect answers.
Instead one should first determine what are the truly physical
degrees of freedom, and subsequently find their interaction
vertices and the ultraviolet cutoff.

As a small sidestep, let us illustrate the above by looking
at an analogy: the simpler and more familiar case of QED.
QED is described by a vector field $A_{\mu}$,
which naively contains 4 dofs. However, one of the 4 dofs
(the longitudinal component of the spatial vector field) is
not physical due to the gauge symmetry
$A_{\mu} \rightarrow A_{\mu}+\partial_{\mu}\Lambda$.
Moreover one of the components of the vector field is
not dynamical: there are no time derivatives acting on $A_0$
in the Maxwell action. This becomes particularly clear
when the theory is written in Hamiltonian form.
The non-dynamical component
(the Coulomb potential) can in fact be decoupled from the
dynamical part of the action. Thus, out of 4 dofs in QED,
there are only 2 remaining dynamical dofs, which correspond to
the transverse components of the vector field, \textit{i.e.}
the two polarizations of the photon.

Let us now turn to Higgs inflation.
The action for Higgs inflation, which basically constitutes of
the Einstein-Hilbert action for general relativity, a scalar
field in a potential and a coupling of the scalar field to
the EH-action, contains naively $10+1$ dynamical degrees of freedom (dofs)
in the metric and scalar field.
However, due to the diffeomorphism symmetry (invariance
under coordinate reparametrizations $x^{\mu}\rightarrow x^{\mu}+\xi^{\mu}$),
4 of these dofs are not physical. The analogy with QED becomes
more apparent when we look at linearized perturbations. The action
for these first order perturbations is invariant under
the transformations $\delta g_{\mu\nu}\rightarrow \delta g_{\mu\nu} +2\nabla_{\left(\mu\right.}\xi_{\left.\nu\right)}$
and $\varphi \rightarrow \varphi +\dot{\phi}\xi^{0}$. This closely
resembles the gauge symmetry of electrodynamics, only in this case
there are 4 gauge parameters, and thus 4 non-physical components
of the metric.

Also, analogous to QED, 4 degrees of freedom
in the metric are not dynamical. These are the $00$ and $0i$
components of the ADM metric,
\begin{equation}
ds^2=-N^2dt^2+g_{ij}(dx^i+N^idt)(dx^j+N^jdt)
\,.
\label{3rdg:ADMlineelement}
\end{equation}
Here $g_{ij}$ is the spatial metric and $N$ and $N^{i}$
are the lapse and shift functions, respectively.
In terms of the ADM metric the action \eqref{Jordanframeaction}
can be written as
\begin{align}
S=\frac12 & \int d^3xdt \sqrt{g}\Biggl\{
N R F(\Phi)+\frac{1}{N}\left(E^{ij}E_{ij}-E^2\right)F(\Phi)
-\frac{2}{N} E F'(\Phi)\left(\partial_t{\Phi}
- N^{i}\partial_i\Phi\right)
\nonumber\\
&+2g^{ij}\nabla_iN \nabla_jF(\Phi)
+\frac{1}{N}\left(\partial_t{\Phi}-N^{i}\partial_i\Phi\right)^2
-Ng^{ij}\partial_i\Phi\partial_j\Phi-2 N V(\Phi)\Biggr\}
\,,
\label{ADMactionnonminimalEij}
\end{align}
where $F'(\Phi)=dF(\Phi)/d\Phi$ and
the measure $\sqrt{g}$, the Ricci scalar $R$
and covariant derivatives $\nabla_i$ are composed
of the spatial part of the metric $g_{ij}$ alone.
The quantities $E_{ij}$ and $E$ are related
to the extrinsic curvature $K_{ij}$ as $E_{ij}=-NK_{ij}$,
with
\begin{align}
E_{ij}&=\frac{1}{2}\left(\partial_tg_{ij}-\nabla_iN_j-\nabla_jN_i\right)\\
E&=g^{ij}E_{ij}
\,.
\end{align}
From Eq.~\eqref{ADMactionnonminimalEij} it can clearly be seen
that the lapse and shift functions are not dynamical (there are no kinetic terms
for them). This becomes even clearer in a Hamiltonian formulation~\cite{Weenink:2010rr}, 
in which the lapse and shift functions
multiply the energy and momentum constraints. Because the lapse and shift functions
are related to the constraints in general relativity and are non-dynamical,
they are also called constraint, or auxiliary fields. Thus, out of the
$10+1$ degrees of freedom (dofs) in the action \eqref{Jordanframeaction}, 4 are non-dynamical
and related to the constraints, and 4 others are gauge dofs. We can therefore
already argue that there are only 3 dynamical dofs in the perturbed action,
and we should take this into account when computing the physical cutoff.

As a first example of how incorrect results are obtained
by not taking into account the gauge freedom and
constraints in general relativity, let us consider
the case of quantum corrections in Higgs inflation
by expanding around a Minkowsky background, as done
in Refs.~\cite{Burgess:2009ea,Barbon:2009ya}. The dominant
contribution from the term $\xi \Phi^2 R$ is found to be
$(\xi/M_P) \varphi^2 \Box h$. However, it is well known
that the only dynamical dofs in Minkowsky space are
the transverse traceless part of the metric, the graviton $\gamma_{ij}$, and
the scalar field fluctuation $\varphi$~~\footnote{Commonly, in Minkowski space the $00$ and $0i$
components of the metric fluctuation $h_{\mu\nu}$ are zero
(the solution for the auxiliary fields is zero), and -- in the absence of matter fields
-- the gauge freedom
is used to set the scalar and vectorial components of the spatial metric
fluctuation $h_{ij}$ to zero. The only remaining degree of
freedom is the traceless and traceless $\gamma_{ij}$.}. In that case,
a term such as $h = \rm{Tr}[h_{\mu\nu}]$ disappears from
the action. The next most dominant term is then
$(\xi/M_P^2) \varphi^2 \partial \gamma_{ij}\partial \gamma_{ij}$,
which is a dimension 6 term with a cutoff scale $M_P/\sqrt{\xi}$,
which is higher than the naive cutoff scale $M_P/\xi$.
Although this is still an incorrect derivation of the cutoff
-- one should perform an expansion around an expanding universe
and a time dependent background field -- it already shows that
taking into account the gauge symmetry can raise the cutoff scale.

So let us continue by considering fluctuations on top of an expanding
universe. In order to deal with the non-dynamical parts of the
metric and the gauge freedom, it is convenient to separate the
metric into different components and expand each component separately.
We thus insert into the action \eqref{Jordanframeaction} (or Eq. \eqref{ADMactionnonminimalEij})
\begin{align}
g_{ij}&=a^2(t){\rm e}^{2\zeta}({\rm e}^\alpha)_{ij}
\nonumber\\
\Phi&=\phi+\varphi
\nonumber\\
N&=\bar{N}\left(1+n\right)
\nonumber\\
N^i&=a^{-1}\bar{N}(t)(a^{-1}\partial_i s+n_{i}^{T})
\label{perturbations}
\,,
\end{align}
where $a(t)$ is the scale factor and
\begin{equation}
\alpha_{ij}=a^{-2}\partial_i\partial_j \tilde h + a^{-1}\partial_{(i}h_{j)}^T+\gamma_{ij}
\,,
\end{equation}
with $\gamma_{ii}=0$, $\partial_j\gamma_{ji}=0$, $\partial_j h^T_{j}=0$ and $\partial_i n_i^T=0$.
Now, all fluctuations of the metric and scalar field
are by themselves not invariant under the gauge transformations (coordinate
reparametrizations) of general relativity. Thus interaction
vertices for the fluctuations in Eq. \eqref{perturbations} are
generally not gauge invariant, and thus not physical. In
order to find the physical interaction vertices, the best way
to go is to derive the action for \textit{gauge invariant perturbations}.
These are variables that are by themselves invariant
under coordinate reparametrizations, and hence their
interaction vertices are physical.
In Refs.~\cite{Weenink:2010rr,Prokopec:2012ug,Prokopec:2013zya} we have precisely done this:
we have derived the quadratic and cubic actions for gauge
invariant cosmological perturbations. We have found that,
when written in terms of physical (gauge invariant) fields,
the scalar-tensor theory~(\ref{Jordanframeaction})
can be separated into dynamical and constraint parts, such
that the constraint and dynamical fields
decouple.
The dynamical part contains one real scalar degree of freedom
and one (traceless, transverse) tensor field with two degrees of freedom
(polarizations).
The non-dynamical part contains gauge invariant versions
of the constraint fields in the action.
There is one gauge invariant scalar lapse function
and one gauge invariant shift vector field (with three components),
which is usually split into a transverse vector and a (longitudinal) scalar.
On-shell all four gauge invariant constraint fields are zero
(zero gauge invariant constraint fields solve the corresponding equations of motion).
This can also be understood by viewing the lapse and shift fields
as auxiliary fields that can be solved for, order by order in perturbation
theory, and their solution inserted in the action \cite{Prokopec:2013zya}
(see also Ref.~\cite{Maldacena:2002vr} for a gauge fixed approach). The
gauge invariant lapse and shift functions being zero on-shell
corresponds to solving the equation of motion for the lapse
and shift function. Solving for the auxiliary fields is hence analogous
to decoupling the fields in the action.
Because of this decoupling, the considerations of the gauge invariant
scalar-tensor theories significantly simplify, in the sense that
it suffices to consider the action for the dynamical
degrees of freedom only (truncated to a certain order).
In Refs.~\cite{Weenink:2010rr,Prokopec:2012ug,Prokopec:2013zya}.
such an action was constructed up to cubic order in the fields
for a class of scalar-tensor theories with general $F(\Phi)$ term
in Eq.~(\ref{Jordanframeaction}).
In order to do that, we had to make a choice for gauge invariant variables.
The potential complication is the fact that, at second order in gauge
transformations there are infinitely many such variables.
However, one can show that
the form of the cubic actions, when expressed in terms of different
gauge invariant variables, differs only by surface
terms~\cite{Prokopec:2012ug,Prokopec:2013zya}, which are irrelevant for
the (bulk) evolution. Nevertheless, these surface terms can be of physical relevance
for (the non-Gaussian part of) the initial state. One set of gauge invariant
variables is special, in that the variables are also frame independent:
they are invariant under a transformation from Jordan to Einstein frame.
When expressed in terms of these variables,
the cubic actions in the Einstein and Jordan frame are
manifestly {\it equal}
(when one takes account of the trivial frame transformation of the time dependent background
quantities that appear as prefactors in front of the cubic vertices).

Therefore, in order to simplify the unitarity cutoff discussion
it is natural to work with these unique frame (and gauge) independent
variables. They are the gauge invariant scalar
pertubation $W_\zeta$ and the gauge invariant tensor (transverse, traceless graviton) perturbation
$\tilde\gamma_{ij}$. These are the second order generalizations of
the gauge invariant Sasaki-Mukhanov field, $w_\zeta=\zeta-(H/\dot\phi)\varphi$,
and the graviton perturbation $\gamma_{ij}$,
where $H(t)=\dot a/a=\bar N(t)^{-1}d\ln(a)/dt$ is the Hubble parameter,
and $\phi(t)$ the background (inflaton) field.
For the precise definitions of $W_\zeta$ and $\tilde\gamma_{ij}$ -- which are given by
$w_\zeta$ and $\gamma_{ij}$ plus corrections that are quadratic in perturbations --
we refer to Ref.~\cite{Prokopec:2013zya}. We have also
given an explicit proof that these variables are frame independent,
thus $W_{\zeta,E}=W_{\zeta}$ and $\tilde{\gamma}_{ij,E}=\tilde{\gamma}_{ij}$.

The complete action for these gauge and frame independent
dynamical fields up to the cubic order
for these variable can be found in
Refs.~\cite{Weenink:2010rr,Prokopec:2012ug,Prokopec:2013zya},
which we shall repeat here.
The quadratic action is a sum of the scalar and tensor parts,
\begin{eqnarray}
S^{(2)}= S^{(2)}_{W_{\zeta}} + S^{(2)}_{\tilde\gamma}
\,,
\label{quadratic action:total}
\end{eqnarray}
where in the Jordan frame (see~(3.23) of Ref.~\cite{Prokopec:2013zya}),
\begin{eqnarray}
S^{(2)}_{W_{\zeta}} &=& \int d^{3}xdt \bar{N} a^{3}
z^2\left[\frac12\dot{W}_{\zeta}^2-\frac12\left(\frac{\partial_i W_{\zeta}}{a}\right)^2\right]
\,,
\label{nmaction:2:wzetawzeta:GI}\\
S^{(2)}_{{\tilde\gamma}_\zeta} &=& \int d^3xdt \bar{N} a^3\frac{F}{4}
\left[\frac12(\dot{\tilde\gamma}_{\zeta ij})^2
-\frac12\left(\frac{\partial_l \tilde\gamma_{\zeta\,ij}}{a}\right)^2\right]
\,,
\label{nmaction:2:gammagamma}
\end{eqnarray}
where $\bar N(t)$ is the background shift function
(for conformal time $t\rightarrow \eta$, $\bar N = a(\eta)$,
while for physical time $t$, $\bar N(t)= 1$).
Note that the prefactor in the integrand of Eq.~(\ref{nmaction:2:wzetawzeta:GI})
is in fact
\begin{equation}
z^2 = \frac{\dot{\phi}^2+\frac{3}{2}\frac{\dot{F}^2}{F}}
{\big(H+\frac12\frac{\dot{F}}{F}\big)^2}
\,.
\label{z:Jordan}
\end{equation}
The cubic action can be conveniently split into the scalar-scalar-scalar,
scalar-scalar-tensor, scalar-tensor-tensor and tensor-tensor-tensor vertices,
\begin{eqnarray}
S^{(3)}= S^{(3)}_{W_{\zeta}W_{\zeta}W_{\zeta}} + S^{(3)}_{W_{\zeta}W_{\zeta}\tilde\gamma}
     + S^{(3)}_{W_{\zeta}\tilde\gamma\tilde\gamma} + S^{(3)}_{\tilde\gamma\tilde\gamma\tilde\gamma}
\,.
\label{cubic action:total}
\end{eqnarray}
Up to boundary terms,
the scalar-scalar-scalar part of the action is (see Ref.~\cite{Prokopec:2012ug}
and Eq.~(6.70) of Ref.~\cite{Weenink:2013oqa})
\begin{eqnarray}
S^{(3)}_{W_{\zeta}W_{\zeta}W_{\zeta}}
 &=& \int d^3x dt\bar{N} a^3 F \Biggl\{
\frac{1}{4}\frac{z^4}{F^2}\left[W_{\zeta}\left(\dot{W}_{\zeta}^2
+\left(\frac{\partial_i W_{\zeta}}{a}\right)^2\right)
-2\dot{W}_{\zeta}\left(\frac{\partial_i}{\nabla^2}
\dot{W}_{\zeta}\right)\partial_i W_{\zeta}\right]
\nonumber\\
&&-\,\frac{1}{16}\frac{z^6}{F^3}W_{\zeta}\left[\dot{W}_{\zeta}^2
-\left(\frac{\partial_i\partial_j}{\nabla^2}\dot{W}_{\zeta}\right)
\left(\frac{\partial_i\partial_j}{\nabla^2}\dot{W}_{\zeta}\right)\right]
+\frac12 \frac{z^2}{F}\left[\frac{\frac{\dot{z}}{z}-\frac12\frac{\dot{F}}{F}}
{H+\frac12\frac{\dot{F}}{F}}\right]^{\cdot}
 \dot{W}_{\zeta}W_{\zeta}^2
\Biggr\}
\,,
\label{3rd:Jordan: Cubic GI Action Wzeta}
\end{eqnarray}
the scalar-scalar-tensor part of the action is (see Ref.~\cite{Prokopec:2013zya}
and Eqs.~(7.12) and~(7.47)  of~\cite{Weenink:2013oqa}),
\begin{eqnarray}
S^{(3)}_{W_{\zeta}W_{\zeta}\tilde{\gamma}_{\zeta}}
&=&\int d^{3}xdt \bar{N} a^{3}
\Biggl\{
\frac12z^2\tilde\gamma_{\zeta,ij}\bigg(\frac{\partial_i}{a}W_\zeta\bigg)
                            \bigg(\frac{\partial_j}{a}W_\zeta\bigg)
-\frac{z^4}{8F}W_\zeta\dot{\tilde\gamma}_{\zeta,ij}\frac{\partial_i\partial_j}{\nabla^2}{\dot W}_\zeta
\nonumber\\
&&\hskip 2.3cm
+\,\frac{z^4}{8F}\bigg(\frac{\partial_i\partial_j}{\nabla^2}{\dot W}_\zeta\bigg)
\bigg(\frac{\partial_k}{\nabla^2}\dot W_\zeta\bigg)
\partial_k\tilde\gamma_{\zeta,ij}
\Biggr\}
\,,\qquad
\label{3rdg:nmaction:3:ssg:GI:Wzeta}
\end{eqnarray}
the scalar-tensor-tensor part of the action is (see Ref.~\cite{Prokopec:2013zya}
and Eq.~(7.49) of~\cite{Weenink:2013oqa}),
\begin{eqnarray}
S^{(3)}_{W_{\zeta}\tilde{\gamma}_{\zeta}\tilde{\gamma}_{\zeta}}
&=&\int d^{3}xdt \bar{N} a^{3}
\Biggl\{
\frac{z^2}{16}W_\zeta\bigg[\dot{\tilde\gamma}_{\zeta,ij}\dot{\tilde\gamma}_{\zeta,ij}
   +\bigg(\frac{\partial_l}{a}\tilde\gamma_{\zeta,ij}\bigg)
   \bigg(\frac{\partial_l}{a}\tilde\gamma_{\zeta,ij}\bigg)\bigg]
-\frac{z^2}{8}\dot{\tilde\gamma}_{\zeta,ij}\bigg(\partial_l\tilde\gamma_{\zeta,ij}\bigg)
    \frac{\partial_l}{\nabla^2}\dot W_\zeta
\Biggr\}
\,,\qquad
\label{3rdg:nmaction:3:sgg:GI:Wzeta}
\end{eqnarray}
and finally the tensor-tensor-tensor part of the action is (see Ref.~\cite{Prokopec:2013zya}
and Eq.~(7.39) of~\cite{Weenink:2013oqa}),
\begin{eqnarray}
S^{(3)}_{\tilde{\gamma}_{\zeta}\tilde{\gamma}_{\zeta}\tilde{\gamma}_{\zeta}}
&=&\int d^3xdt \bar{N} a^3 \frac{F}{8}
\Biggl\{
\tilde{\gamma}_{\zeta,ij}
\frac{\partial_i\tilde{\gamma}_{\zeta,kl}}{a}
\frac{\partial_j\tilde{\gamma}_{\zeta,kl}}{a}
+\tilde{\gamma}_{\zeta,kl}
\frac{\partial_i\tilde{\gamma}_{\zeta,kj}}{a}
\frac{\partial_j\tilde{\gamma}_{\zeta,il}}{a}
-\tilde{\gamma}_{\zeta,ik}
\frac{\partial_i\tilde{\gamma}_{\zeta,jl}}{a}
\frac{\partial_j\tilde{\gamma}_{\zeta,kl}}{a}
\Biggr\}
\,.\qquad
\label{3rdg:nmaction:3:ggg:GI:Wzeta}
\end{eqnarray}

When written in terms of the frame independent variables $W_\zeta$ and $\tilde\gamma_{\zeta ij}$,
we proved in~\cite{Prokopec:2013zya}
that the action~(\ref{quadratic action:total}--\ref{3rdg:nmaction:3:ggg:GI:Wzeta})
is frame independent.
Since the frame independent perturbations themselves coincide in Jordan
and Einstein frame, $W_\zeta=W_{\zeta,E}$ and $\tilde\gamma_{\zeta ij}=\tilde\gamma_{\zeta ij,E}$,
the Einstein frame and Jordan frame actions coincide as well.
The frame independence becomes manifest when
one expresses the background quantities in the Jordan frame $a,H,z$ in
terms of their Einstein frame counterparts
in Eqs.~(\ref{quadratic action:total}--\ref{3rdg:nmaction:3:ggg:GI:Wzeta})
by using the relations \eqref{Conformaltransformationmetricscalar}
at background level. This gives the following identities
\begin{eqnarray}
F^{1/2}\bar N &=& M_{\rm P}\bar{N}_E
\,,\quad
F^{1/2}a= M_{\rm P}a_E
\,,\quad
F^{-1/2}\dot W_\zeta  = M_{\rm P}^{-1}\dot W_{\zeta,E}
\,,\quad
\frac{H+\frac{\dot F}{2F}}{\sqrt{F}}= \frac{H_E}{M_{\rm P}}
\nonumber\\
\frac{\dot\phi^2+\frac32\frac{\dot F^2}{F}}{F^2}&=& \frac{\dot{\phi}_E^2}{M_{\rm P}^4}
\,,\qquad
\frac{z}{\sqrt{F}}= \frac{z_E}{M_{\rm P}}
\,,\qquad
\frac{\frac{\dot z}{z}-\frac12\frac{\dot F}{F}}{H+\frac{\dot F}{2F}}
                 = \frac{1}{M_{\rm P}}\frac{\dot{z}_E}{z_E}
\,,
\label{from Jordan to Einstein}
\end{eqnarray}
where all subscripts $E$ denote quantities in the Einstein frame, $z_E\equiv\dot{\phi}_E/H_E$
and a dotted derivative denotes $\dot{X}=dX/(\bar{N}dt)$ and $\dot{X}_E=dX_E/(\bar{N}_Edt)$
for Jordan and Einstein frame quantities respectively.
Hence there is no need to quote the action in the Einstein frame
(in which $z^2/F=z_E^2/M_P^2=\dot{\phi}_E^2/(M_P^2H_E^2)$).

In order to compute the cutoff in a frame independent formulation,
it is convenient to transform to canonically normalized variables
\begin{equation}
V_\zeta  =  a z W_{\zeta} = a \sqrt{2\epsilon_E F } W_{\zeta}
\,,\qquad
\Gamma_{\zeta,ij}  =  \frac12 a \sqrt{F} \tilde{\gamma}_{\zeta,ij}
\,,
\label{3rdg:canonicalvariables}
\end{equation}
where we have used
\begin{equation}
\epsilon_E = -\frac{\dot{H}_E}{H_E^2} = \frac{z_E^2}{2M_P^2} = \frac{z^2}{2F}
\,,
\end{equation}
which relates $z$ (defined in Eq.~\eqref{z:Jordan})
to the slow-roll parameter $\epsilon_E$ in the Einstein frame.
In terms of the canonical fields the
second order action~(\ref{quadratic action:total}--\ref{nmaction:2:gammagamma})
in conformal time $\tau$ ($\bar{N}(t)\rightarrow a(\tau)$) become
\begin{eqnarray}
S^{(2)}_{V_\zeta}&=&\frac12\int d^3xd\tau
\left[V_\zeta^{\prime 2}-(\partial_i V_\zeta)^2+\frac{(az)^{\prime\prime}}{az}V_\zeta\right]
-\frac12\int d^3 x \bigg[\frac{(az)^{\prime}}{az}V_\zeta^2
                   \bigg]_{\tau_{\rm in}}^{\tau_{\rm fin}}
\label{scalar quadratic action:canonical}\\
S^{(2)}_{\Gamma_{\zeta,ij}}&=&\frac12\int d^3x d\tau
\left[\Gamma_{\zeta,ij}^{\prime 2}-(\partial_l \Gamma_{\zeta,ij})^2
        +\frac{(a\sqrt{F})^{\prime\prime}}{a\sqrt{F}}\Gamma_{\zeta,ij}^2\right]
-\frac12\int d^3 x \bigg[\frac{(a\sqrt{F})^{\prime}}{a\sqrt{F}}\Gamma_{\zeta,ij}^2
                   \bigg]_{\tau_{\rm in}}^{\tau_{\rm fin}}
\,,
\label{tensor quadratic action:canonical}
\end{eqnarray}
where a prime denotes a derivative with respect to conformal
time $\tau$. The boundary terms
in~(\ref{scalar quadratic action:canonical}--\ref{tensor quadratic action:canonical})
do not contribute to the propagator equation of motion and thus they can be discarded.
Furthermore, the transformation to the canonically normalized fields has led to generation of
time dependent (negative) mass terms
in~(\ref{scalar quadratic action:canonical}--\ref{tensor quadratic action:canonical}).
However, these terms can be neglected on energy and momentum scales far above the Hubble scale.
Namely, $(az)^\prime/(az)=aH+(1/2)\phi^\prime [d\ln(F)/d\phi]+(1/2)[\epsilon_E^\prime/\epsilon_E]$.
Now, since the latter two terms are suppressed by slow roll parameters,
the first term, $aH=\cal{H}$ (here and subsequently ${\cal H}$ denotes a conformal
Hubble rate) is the dominant term. Likewise, one can show that the dominant term
in $(az)^{\prime\prime}/(az) = 2a^2H^2$ plus slow roll suppressed terms, such that when
the (conformal) energy scale, $E_c\gg {\cal H}$, this term can be neglected.
Because $(a\sqrt{F})^{\prime\prime}/(a\sqrt{F})= 2a^2 H^2$ plus slow roll suppressed terms,
the same consideration applies to canonically normalized gravitons.
This means that the canonically normalized scalar and graviton propagator behave in the
ultraviolet (on scales $E_c = |k^0|\gg {\cal H}$ and $\|\vec k\|\gg {\cal H}$) as,
$\sim 1/(\eta_{\mu\nu}k^\mu k^\nu)$ plus corrections of the order ${\cal H}^2$, which
we shall neglect in the following considerations.

 When the cubic action~(\ref{3rd:Jordan: Cubic GI Action Wzeta}--\ref{3rdg:nmaction:3:ggg:GI:Wzeta})
is expressed in terms of the canonical variables~(\ref{3rdg:canonicalvariables}),
one gets for the  pure scalar cubic action
%
\begin{eqnarray}
S^{(3)}_{V_{\zeta}V_{\zeta}V_{\zeta}} &\simeq& \int d^3xd\tau \sqrt{\frac{\epsilon_E}{8a^2F}}
\bigg\{
  V_\zeta\bigg[
            (V_\zeta^\prime)^2-2\frac{(az)^\prime}{az}V_\zeta^\prime V_\zeta
             +\frac{[(az)^\prime]^2}{(az)^2}V_\zeta^2 +(\partial_iV_\zeta)^2
         \bigg]
\label{pure scalar cubic:canonical}
\\
   && \hskip 2.9cm  -\, 2\bigg(V_\zeta^\prime-\frac{(az)^\prime}{az}V_\zeta\bigg)
 \bigg[\frac{\partial_i}{\nabla^2}\bigg(V_\zeta^\prime-\frac{(az)^\prime}{az}V_\zeta\bigg)\bigg]
          (\partial_iV_\zeta)
\nonumber\\
   && \hskip 2.9cm
 -\,\frac{\epsilon_E}{2}V_\zeta\bigg[
            \bigg(\!V_\zeta^\prime\!-\frac{(az)^\prime}{az}V_\zeta\bigg)^2
\!-\frac{\partial_i\partial_j}{\nabla^2}\bigg(V_\zeta^\prime\!-\!\frac{(az)^\prime}{az}V_\zeta\bigg)
-\frac{\partial_i\partial_j}{\nabla^2}\bigg(V_\zeta^\prime\!-\!\frac{(az)^\prime}{az}V_\zeta\bigg)
         \bigg]
\nonumber\\
   && \hskip 2.9cm
+\frac{1}{\epsilon_E}\bigg(
                         \frac{\epsilon_E^\prime/\epsilon_E}{2{\cal H}+(\partial_\tau F)/F}
                     \bigg)^\prime
                  V_\zeta^2\bigg(\!V_\zeta^\prime\!-\frac{(az)^\prime}{az}V_\zeta\bigg)
\bigg\}
\,.
\nonumber
\end{eqnarray}
Next, the scalar-scalar-tensor cubic action~(\ref{3rdg:nmaction:3:ssg:GI:Wzeta})
can be written as,
\begin{eqnarray}
S^{(3)}_{V_{\zeta}V_{\zeta}\Gamma_\zeta}
&=&\int d^{3}xd\tau
\Biggl\{
\frac{1}{a\sqrt{F}}\Gamma_{\zeta,ij}\big(\partial_iV_\zeta\big)\big(\partial_jV_\zeta\big)
\nonumber\\
&&\hskip 1.5cm
-\,\frac{z^2}{4aF^{3/2}}V_\zeta\Bigg(\Gamma_{\zeta,ij}^\prime-\frac{(a\sqrt{F})^\prime}{a\sqrt{F}}\Gamma_{\zeta,ij}\bigg)
    \frac{\partial_i\partial_j}{\nabla^2}\bigg(V_\zeta^\prime-\frac{(az)^\prime}{az}V_\zeta\bigg)
\nonumber\\
&&\hskip 1.5cm
+\,\frac{z^2}{4aF^{3/2}}\frac{\partial_i\partial_j}{\nabla^2}\bigg(V_\zeta^\prime-\frac{(az)^\prime}{az}V_\zeta\bigg)
\frac{\partial_l}{\nabla^2}\bigg(V_\zeta^\prime-\frac{(az)^\prime}{az}V_\zeta\bigg)
\partial_l\Gamma_{\zeta,ij}
\Biggr\}
\,,\qquad
\label{dominant ssg}
\end{eqnarray}
the scalar-tensor-tensor part of the action~(\ref{3rdg:nmaction:3:sgg:GI:Wzeta}) becomes,
\begin{eqnarray}
S^{(3)}_{V_{\zeta}\Gamma_{\zeta}\Gamma_{\zeta}}
&=&\int d^{3}xd\tau
\Biggl\{
\frac{z}{4aF}V_\zeta\bigg[\bigg(\Gamma_{\zeta,ij}^\prime
                       -\frac{(a\sqrt{F})^\prime}{a\sqrt{F}}\Gamma_{\zeta,ij}\bigg)
    \bigg(\Gamma_{\zeta,ij}^\prime-\frac{(a\sqrt{F})^\prime}{a\sqrt{F}}\Gamma_{\zeta,ij}\bigg)
   +\big(\partial_l\Gamma_{\zeta,ij}\big)\big(\partial_l\Gamma_{\zeta,ij}\big)\bigg]
\nonumber\\
&&\hskip 1.5cm
-\,\frac{z}{2aF}\bigg(\Gamma_{\zeta,ij}^\prime
                    -\frac{(a\sqrt{F})^\prime}{a\sqrt{F}}\Gamma_{\zeta,ij}\bigg)
   \big(\partial_l\Gamma_{\zeta,ij}\big)
    \frac{\partial_l}{\nabla^2}\bigg(V_\zeta^\prime-\frac{(az)^\prime}{az}V_\zeta\bigg)
\Biggr\}
\,,\qquad
\label{dominant sgg}
\end{eqnarray}
and finally the pure tensor part of the action~(\ref{3rdg:nmaction:3:ggg:GI:Wzeta}) becomes
\begin{eqnarray}
S^{(3)}_{\Gamma_{\zeta}\Gamma_{\zeta}\Gamma_{\zeta}}
\!=\!\int\! \frac{d^3xd\tau}{a\sqrt{F}}
\biggl\{
\Gamma_{\zeta,ij}\big(\partial_i\Gamma_{\zeta,kl}\big)\big(\partial_j\Gamma_{\zeta,kl}\big)
\!+\!\Gamma_{\zeta,kl}\big(\partial_i\Gamma_{\zeta,kj}\big)\big(\partial_j\Gamma_{\zeta,il}\big)
\!-\!\Gamma_{\zeta,ik}\big(\partial_i\Gamma_{\zeta,jl}\big)\big(\partial_j\Gamma_{\zeta,kl}\big)
\!\biggr\}
\,.\quad
\label{dominant ggg}
\end{eqnarray}

We shall use the cubic action~(\ref{pure scalar cubic:canonical}--\ref{dominant ggg})
in section~\ref{sec: cutoffscale} to analyse the unitarity cutoff
implied by the $2\to 2$ tree-level scattering processes.
Of course, to complete the analysis,
one would also need the gauge invariant quartic vertices, which are currently not
available. We do not expect however, that quartic vertices
will change in any way the qualitative discussion we present in the next section.

\section{The cutoff scale revisited}
\label{sec: cutoffscale}

A usual requirement for renormalizable theories
is that the unitarity bound is not violated in
perturbation theory, \textit{i.e.} scattering
amplitudes should not become bigger than unity.
In particular there is the requirement of tree unitarity~\cite{Cornwall:1974km},
which states that $N$ particle tree amplitudes
$\mathcal{A}_N$ should not grow more rapidly than
$E^{4-N}$, where $E$ is the center of mass energy.
If the amplitude grows faster, perturbation theory
fails at some cutoff scale $\Lambda$. This usually
means that some new physics should enter at this energy
scale. When the relevant energy scales of the theory
under consideration are well below the cutoff scale,
the theory can be considered 'natural', in the sense
that perturbation theory is valid and there is no
need for new physics. Conversely, there is a naturalness problem
if typical energy scales are higher than the cutoff scale.

The cutoff is most easily computed when the perturbative
action is written in canonical form, such that
the propagator goes as $1/k^2$, where $k^2 = \eta_{\mu\nu}k^\mu k^\nu$
and $k^\mu$ is a conformal 4-momentum.
Next, the cutoff can be read off from the vertices of dimension
higher then 4, which should be suppressed as $\Lambda^{4-D}$.
This has been done for Higgs inflation in Ref.~\cite{}
in both the Einstein and Jordan frame, and we
outline the derivation here. In the Jordan frame the starting point is
the action~\eqref{Jordanframeaction} with $F(\Phi)=M_P^2+\xi\Phi^2$,
which is the action for Higgs inflation in unitary gauge
and gauge interactions are neglected.
Next, following Ref.~\cite{Bezrukov:2011sz},
and analogously as it was done in~(\ref{3rdg:canonicalvariables}),
generic (non-invariant) perturbations $\delta g_{\mu\nu}=g_{\mu\nu}-\bar g_{\mu\nu}$
and $\varphi=\Phi(x)-\phi(t)$ can be rescaled to canonical variables $\delta\hat g_{\mu\nu}$
and $\hat\varphi$ for which the corresponding quadratic actions are canonically normalized.
Here $\bar g_{\mu\nu}$ denotes a background metric (which in the UV can be approximated
by Minkowski metric) and $\phi(t)$ is a background field.
The dominant term in the action with dimension higher than $4$ and
in the Jordan frame is of the order
\begin{equation}
\xi \varphi^2 \Box \delta g
\,,
\label{dominant vertex: dim > 4}
\end{equation}
where $\delta g = \bar{g}^{\mu\nu} \delta g_{\mu\nu}$.
When reexpressed in terms of the canonically normalized fields  $\hat g_{\mu\nu}$
and $\hat\varphi$, this term becomes
\begin{equation}
 \frac{\xi\sqrt{M_P^2+\xi\phi^2}}{M_P^2+\xi\phi^2+6\xi^2\phi^2}
\hat{\varphi}^2 \Box {\delta\hat g}
\,.
\label{dominant vertex: dim > 4:2}
\end{equation}
At high energies this vertex scales as $E^2$, where $E=|k^0|$ denotes the energy scale. If
we consider a $2\rightarrow 2$ scattering process of $\hat{\varphi}$
via exchange of a gravitational scalar $\delta\hat g$, we see that
the total amplitude scales as $E^2/\Lambda^2$ at high energies. The cutoff scale
is precisely the inverse of the operator above. The cutoff in
the Jordan frame is thus
\begin{equation}
\Lambda \sim \frac{M_P^2+\xi\phi^2+6\xi^2\phi^2}{\xi\sqrt{M_P^2+\xi\phi^2}}
\,.
\label{cutoff scale:from paper}
\end{equation}
In Ref.~\cite{Bezrukov:2010jz} the cutoff
was also computed \textit{via} the Einstein frame. In that frame
the non-minimal coupling term is absent and the gravitational
and field kinetic terms are canonical. Still, the cutoff
scale reappears in the non-polynomial potential and shows
a similar  behavior as above (though not exactly equal).
From~(\ref{cutoff scale:from paper}) one sees that
$\Lambda \propto \sqrt{\xi} \phi$ in the regime where $\phi\gg M_P/\sqrt{\xi}$,
$\Lambda \sim \xi \phi^2/M_P$ for $M_P/\sqrt{\xi} \gg \phi \gg M_P/\xi$ and
$\Lambda \sim M_P/\xi$ when $\phi\ll M_P/\xi$. The authors
of Ref. \cite{Bezrukov:2011sz} then argue that all relevant
energy scales in these regimes are lower than the cutoff scale,
such that the perturbative expansion is valid
and Higgs inflation is natural.

These results are interesting, but the question arises whether they can be trusted.
In the following we point at the principal potential problems with
the analysis outlined above, which put into question the reliability of
the cutoff scale in Eq.~(\ref{cutoff scale:from paper})

\begin{itemize}

\item[{\bf a)}] {\bf The computation of the cutoff is gauge dependent.}

Both the metric fluctuations $\delta g_{\mu\nu}$ and scalar field
perturbation $\varphi$ are gauge dependent, hence the
vertex~(\ref{dominant vertex: dim > 4}) is also gauge dependent, and has on
its own no physical meaning. Indeed, it is well known that,
when working with gauge dependent quantities, one can reach realiable conclusions
only when one takes account of all terms up to this order.
Namely, different gauge dependent terms can cancel each other, which
was not accounted for in Ref.~\cite{Bezrukov:2011sz}, see {\it e.g.} Ref.~\cite{George:2013iia}.
Furthermore, when written in terms of gauge invariant variables,
gauge dependent vertices (such as $\varphi^2\Box \delta g$)
can be absorbed into the lower order action, such that they simply `disappear' from
a gauge invariant action. Next, some metric perturbations are non-dynamical
in the sense that they act as auxiliary (constraint) fields and should be solved for,
which generate additional vertices that may cancel
the problematic vertex. In fact, this already
happens at the level of the quadratic action. Na\"\i vely,
the field perturbation has an effective mass term
$m^2_{\rm eff}\varphi^2=(-\xi \bar{R} + V'')\varphi^2+..$,
where $\bar{R}=6(2H^2+\dot{H})$ is the background Ricci scalar. Such a mass term
is huge, in the sense that $|\xi \bar{R}| \gg H^2$.
However, only a light inflaton field, with $m_{\rm eff}^2 \ll H^2$
can generate a nearly scale invariant power spectrum.
Thus, such a mass term is disastrous for the model
and na\"\i vely rules out Higgs inflation. But, when
contributions from the auxiliary fields are taken into
account, the problematic $\xi \bar R$ contribution gets canceled, leaving
only a light effective mass for the inflaton field.
Similar cancellations occur at higher orders in a gauge
invariant formulation.

\item[{\bf b)}] {\bf The canonical redefinition mixes up frames.}

A crucial step in the computation of the cutoff
was the definition of new perturbations $\hat{\varphi}$
and ${\delta \hat g}$ which canonically normalize
the kinetic terms. However, this redefinition
is in fact nothing more than a transformation
from the Jordan to the Einstein frame at the level
of perturbations~\footnote{This can be explicitly checked
by comparing Eqs. (2.9) and (2.10) in Ref.~\cite{Bezrukov:2010jz}
to the transformations
\[
\varphi_E=\frac{d\phi_E}{d\phi}\varphi+{\cal O}(\varphi^2)
\,,\qquad
\zeta_E=\zeta+\frac{1}{2F}\frac{dF}{d\phi}\varphi+{\cal O}(\varphi^2)
\,,\qquad
\gamma_{ij,E}=\gamma_{ij}
\]
and the expansion of $g_{\mu\nu,E}=\Omega^2 g_{\mu\nu}$ to first order
in perturbations.}!
On the other hand, the cutoff is computed
from the term $\xi\Phi^2 R$ in the Jordan frame action. So somehow one
computes a cutoff from a Jordan frame vertex using
Einstein frame perturbations, which is very bizarre.
Moreover, we would like to emphasize that, although
the Einstein frame is often referred to as the frame
in which both the gravitational action and scalar field action
are written in canonical form, \textit{the Einstein frame is not canonical}.
Canonical formulation means that the kinetic sectors for
different fields are decoupled (and canonically normalized), such
that one can straightforwardly extract the canonical
momentum and quantize the theory. However, in the "canonical" Einstein
frame the gravitational field still couples to
the scalar sector as $\sqrt{-g_E}g^{\mu\nu}_E \partial_{\mu}\Phi_E\partial_{\nu}\Phi_E$.
Conversely, the scalar field still couples to the kinetic term for
the metric perturbations in the $\sqrt{-g_E}R_E$ term via
the auxiliary fields in the metric (which contain
$\varphi_E$ in its first order solution). Hence,
the Einstein frame is formulated in a non-canonical
way, just like the Jordan frame. The true
canonical formulation is only reached once one inserts
perturbations and decouples physical and non-physical
degrees of freedom, which was done in Ref.~\cite{Prokopec:2013zya},
and the results of which are summarized in section~\ref{sec: gauge invariance}.
When using the frame independent variables $W_\zeta$ and $\tilde\gamma_{\zeta,ij}$
the Jordan frame quadratic and cubic actions
becomes manifestly equal to the Einstein frame actions, see Eq.~(\ref{from Jordan to Einstein}).
Thus in a frame independent formulation the notion of frames becomes meaningless.

\item[{\bf c)}] {\bf Inequivalence of Jordan and transformed Einstein cutoff.}

As we have mentioned, the cutoffs computed directly
in the Jordan frame, or $\textit{via}$ the Einstein frame,
are similar, but not exactly the same. However, they should
coincide, because no physical content is lost in the frame
transformation. Of course the origin of this problem is related
to the already mentioned problems, namely a non-invariant formulation
and a mixing-up of different frames.

\end{itemize}

\subsection{Frame independent computation of cutoff}
\label{Frame independent computation of cutoff}

 This subsection we show that the above problems become all obsolete
once the theory is written in a manifestly gauge invariant and frame
independent way. Firstly, when one makes use
of gauge invariant variables, all vertices are physical
vertices. Moreover, the gauge invariant perturbations
decouple in the quadratic action, which makes it very
easy to write the theory in canonically normalized form.
And obviously, when frame independent perturbations are
used, results in the Jordan and Einstein frames become manifestly equivalent.

 As mentioned above, a simple way of estimating the unitarity cutoff scale for
tree tree-level scattering amplitudes is to work with the canonically normalized
fields~(\ref{3rdg:canonicalvariables}), for which the propagator in the
ultraviolet acquires a simple form: $\sim (\eta_{\mu\nu}k^\mu k^\nu)^{-1}$
plus corrections that are suppressed by ${\cal H}^2=(aH)^2$,
where $k^\mu=(E_c/c,\vec k\,)$ denotes a conformal energy and momentum
(the corresponding physical energy and momentum are given by $E=E_c/a$ and $\vec k/a$).
Up to boundary terms, the cubic actions for various combinations of scalar and tensor
vertices are given in Eqs.~(\ref{pure scalar cubic:canonical}--\ref{dominant ggg}).

 Let us first consider the scalar cubic action~(\ref{pure scalar cubic:canonical}).
The cubic vertex ${\cal V}$ from the first two lines is of the order,
\begin{equation}
{\rm Scalar\; cubic\; vertex:}\quad
{\cal V}_{V_\zeta^3} \sim \frac{\epsilon_E^{1/2}{\rm max}\big[E_c^2,\|\vec k\|^2\,\big]}{a\sqrt{F}}
\,,
\label{scalar cubic vertex}
\end{equation}
where we made use of $\partial_\tau\rightarrow E_c$,
$\partial_i\rightarrow k^i$ and we took ultraviolet limit in which
$(az)^\prime/(az)\sim {\cal H}\ll E_c, \|\vec k\,\|$.
The terms in the third line in~(\ref{pure scalar cubic:canonical}) lead to a vertex that is
in addition suppressed by $\epsilon_E$, which is during inflation less than unity, and hence
can be neglected. Finally, the vertex contribution from the fourth line is of the order
$\epsilon_E^{-1/2}[\epsilon_E^\prime/(\epsilon_E{\cal H})]^\prime E_c$, that means it is
proportional to a third order slow roll parameter,
$\epsilon^{(3)}=[\epsilon_E^\prime/(\epsilon_E{\cal H})]^\prime/(\epsilon_E{\cal H})$,
and which can be assumed to be small during inflation. More precisely,
this vertex contribution is suppressed with respect to~(\ref{scalar cubic vertex}) if
${\rm min}\big[E_c,\|\vec k^2\|/E_c]\gg [\epsilon_E^\prime/(\epsilon_E{\cal H})]^\prime/\epsilon_E
=\epsilon^{(3)}{\cal H}$, which one can safely assume to be the case throughout inflation.
On the other hand, we know that in four space-time dimensions a 2-to-2 scattering aplitude
scales as,
\begin{equation}
\frac{{\rm max}[E_c^2,\|\vec k\,\|^2]}{\Lambda^2}
           \sim \frac{{\cal V}^2}{{\rm max}[E_c^2,\|\vec k\|^2]}
\,.
\label{general cutoff for 2-2 scattering}
\end{equation}
Upon combining this with~(\ref{scalar cubic vertex}) we finally get for the physical cutoff
in the Jordan frame,
\begin{equation}
{\rm Scalar\; cubic\; vertex:}\quad
\left(\frac{\Lambda}{a}\right)_J \sim \sqrt{\frac{M_{\rm P}^2+\xi\phi^2}{\epsilon_E}}
           \gtrsim \sqrt{M_{\rm P}^2+\xi\phi^2}
\,,
\label{cutoff Jordan: pure scalar}
\end{equation}
where we took account of $F=M_{\rm P}^2+\xi\phi^2$
and $\epsilon_E\lesssim 1$ during inflation.
In conclusion, we have found that during entire Higgs inflation
for scatterings mediated by the scalar cubic interactions the unitarity bound is
above the Planck scale. In fact, the cutoff in Higgs inflation is higher than the
unitarity cuttoff in the minimally coupled inflation, $\sim M_{\rm P}$.
which is obtained by simply setting $\xi\rightarrow 0$ in~(\ref{cutoff Jordan: pure scalar}).

 Now making use of~(\ref{from Jordan to Einstein}),
from which we see that
\begin{equation}
 a_E=a_J \frac{F^{1/2}}{M_{\rm P}}
\,,
\label{scale factor in two frames}
\end{equation}
the cutoff~(\ref{cutoff Jordan: pure scalar}) can be written in the Einstein frame as
\begin{equation}
\bigg(\frac{\Lambda}{a}\bigg)_E \sim \frac{M_{\rm P}}{\epsilon_E} \gtrsim M_{\rm P}
\,.
\label{dominant scalar cubic vertex:4}
\end{equation}
The difference between the frames can be attributed to the frame dependence of the
physical cutoff scale $\Lambda/a$, and has no physical meaning.
Therefore, to make a meaningful comparison of the Einstein and Jordan frame cutoffs,
the rescaling~(\ref{scale factor in two frames}) has to be taken account of.
Hence, we conclude that there is a perfect agreement in the cutoff scale in two different frames,
and in both frames the physical cutoff is above the Planck scale $M_{\rm P}= 1/(8\pi G_N)^{1/2}$.

Let us now turn our attention to other cubic vertices.
From Eqs.~(\ref{dominant ssg}--\ref{dominant ggg}))
we obtain the dominant contributions for the other types of vertices,
\begin{eqnarray}
&{\rm Scalar-scalar-tensor\; vertex:}\quad &
{\cal V}_{V_\zeta^2\Gamma_\zeta}\sim
            \frac{1}{aF^{1/2}}\times{\rm max}\Big[\epsilon_EE_c^2, \|\vec k\,\|^2\Big]
\nonumber\\
&{\rm Scalar-tensor-tensor\; vertex:}\quad &
{\cal V}_{V_\zeta\Gamma_\zeta^2}\sim
          \frac{\epsilon_E^{1/2}}{aF^{1/2}}\times{\rm max}\Big[E_c^2, \|\vec k\,\|^2\Big]
\nonumber\\
&{\rm Pure\; tensor\; vertex:} \qquad\qquad\quad\quad\;&
{\cal V}_{\Gamma_\zeta^3}\quad\sim \frac{\|\vec k\,\|^2}{aF^{1/2}}
\label{dominant other cubic vertices}
\end{eqnarray}
When these are inserted into~(\ref{general cutoff for 2-2 scattering})
one obtains the following results for the cutoff scales from different types of vertices,
\begin{eqnarray}
&{\rm Scalar-scalar-tensor\; vertex:}\quad &
\left(\frac{\Lambda}{a}\right)_J
\sim \sqrt{M_{\rm P}+\xi\phi^2}\times{\rm min}\Big[\epsilon_E^{-2},1\Big]
    \gtrsim \sqrt{M_{\rm P}+\xi\phi^2}
\nonumber\\
&{\rm Scalar-tensor-tensor\; vertex:}\quad &
\left(\frac{\Lambda}{a}\right)_J \sim \sqrt{M_{\rm P}+\xi\phi^2}
        \times{\rm min}\Big[\epsilon_E^{-1},1\Big]\gtrsim \sqrt{M_{\rm P}+\xi\phi^2}
\label{cutoffs individual vertices}
\\
&{\rm Pure\; tensor\; vertex:} \qquad\qquad\quad\quad\;&
\left(\frac{\Lambda}{a}\right)_J \sim \sqrt{M_{\rm P}+\xi\phi^2}
       \times{\rm min}\Big[E_c^2/\|\vec k\|^2,1\Big]\gtrsim \sqrt{M_{\rm P}+\xi\phi^2}
\,,\qquad
\nonumber
\end{eqnarray}
where the first expression in square brackets gives a cutoff for the
case when $E_c\gg \|\vec k\|$, and the latter for the case when  $E_c\ll \|\vec k\|$.
Analogous conclusions as in~(\ref{cutoffs individual vertices})
are reached when one considers $2-to-2$ scatterings composed by two classes of 
vertices, {\it e.g.} a combination of a scalar-scalar-graviton and a pure graviton vertex.

In summary, we have shown that in all cases, for all kinds of vertices and throughout inflation
the physical cutoff in the Jordan frame is
above the scale $\sqrt{M_{\rm P}+\xi\phi^2}$, while in the Einstein frame it is
above $M_{\rm P}$ (the difference has no physical meaning and it is attributed to the
non-invariant definition of the physical cutoff in the two frames)
\footnote{It comes as no surprise that the canonically normalized
scalar-scalar-graviton and pure graviton vertices are not suppressed
by a factor of $\sqrt{\epsilon_E}$. The reason is that in the
de Sitter limit $\epsilon_E \rightarrow 0$ the curvature perturbation $\zeta$
becomes a pure gauge mode, and is completely absorbed by the gauge invariant
lapse and shift perturbations. The only remaining dynamical perturbations
are the scalar $\varphi$ and the graviton $\gamma_{ij}$.
The term $g^{\mu\nu}\partial_{\mu}\Phi\partial_{\nu}\Phi$ in the original action
then gives the interaction term
$\gamma_{ij}\partial_i\varphi\partial_j\varphi$,
which is not $\epsilon_E$ suppressed.
Likewise, the pure graviton vertices are always present
and are not suppressed by powers of $\epsilon_E$.}.

\begin{figure}[ttt]
\begin{center}
\includegraphics[width=0.8\textwidth]{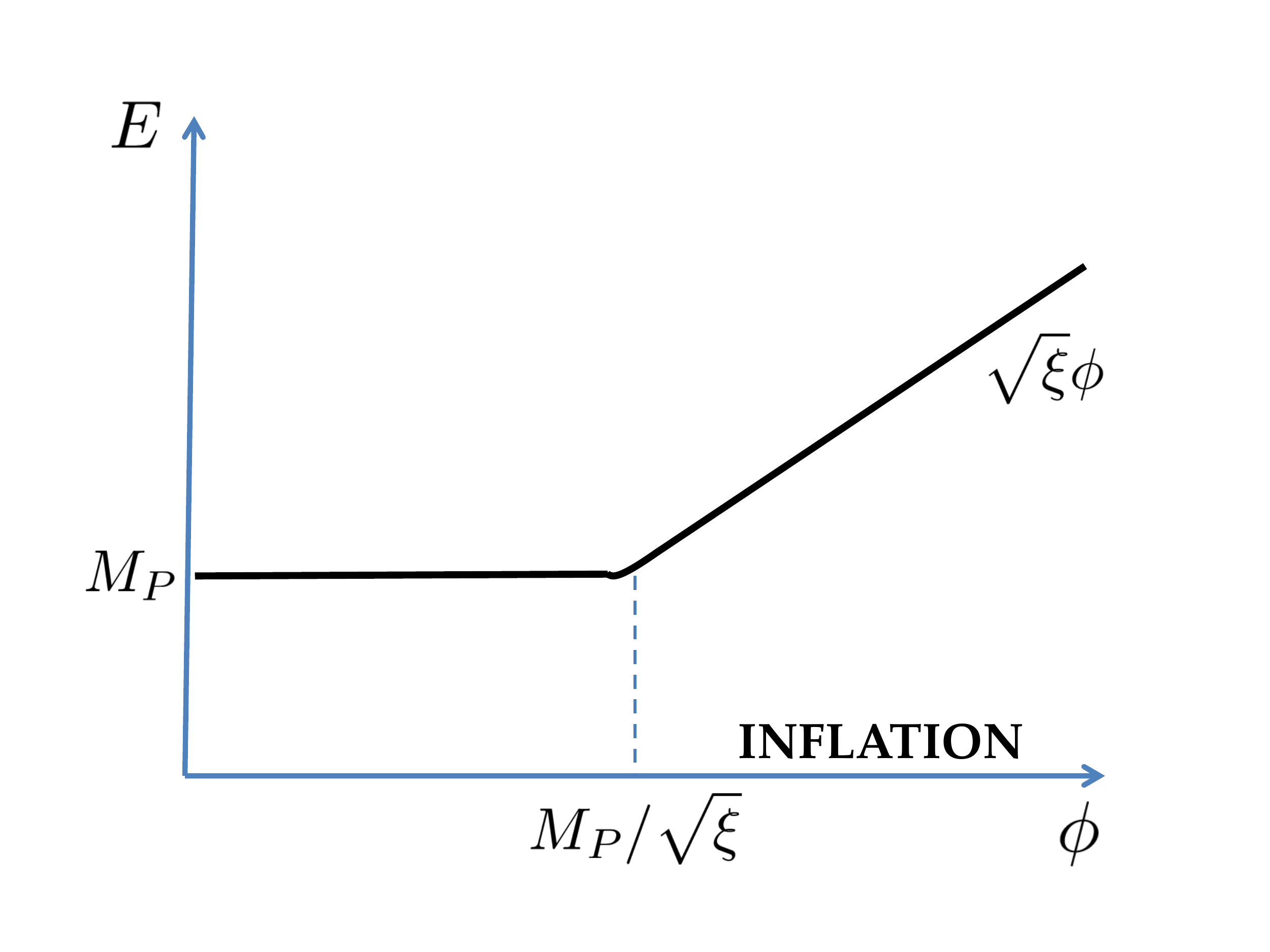}\\
  \caption{Physical cutoff $(\Lambda/a)_J=\sqrt{M_P^2+\xi\phi^2}$ in the Jordan frame.}
\label{fig:cutoffJordan}
\end{center}
\end{figure}
The Jordan frame cutoff is shown in figure
\ref{fig:cutoffJordan}.
The important point is that also
here the cutoff is never smaller than
the Planck scale. This means that
the perturbative expansion is valid
at least up to an energy scale of the order the Planck
scale for a theory with some non-minimal coupling to gravity, such as Higgs
inflation. Thus, the above analysis strongly suggests that,
at least for the class of tree level $2\rightarrow 2$ scattering
processes considered here, there is no naturalness problem in Higgs inflation.

There are caveats to this statement however. Firstly,
we have only considered the scalar-gravity sector
of Higgs inflation, but neglected interactions
with \textit{e.g.} gauge fields. In Ref.~\cite{Burgess:2010zq} it was stated that
the (low) cutoff scale of $M_P/\xi$ also appears in
the Higgs-gauge interactions. However, these vertices
are gauge dependent as well, so the problem may be absent in a gauge invariant
formulation that includes the gauge fields.
Secondly, for the computation of the cutoff
we have used the partially integrated
actions~(\ref{pure scalar cubic:canonical}--\ref{dominant ggg}).
Had we not made use of partial integrations, we would
have found disastrous terms in the action.
For example, before any partial integration, the leading term in
the pure cubic scalar action for $V_\zeta=W_{\zeta}/(az)$ from Ref.~\cite{Prokopec:2013zya}
contributes (in the limit $E_c\gg \|\vec k\|$) to the cubic vertex as,
$(\epsilon_E F/a)\times{\rm max}[E_c^2,\|\vec k\|^2]$, implying a physical
cutoff (in the Jordan frame),
$(\Lambda/a)_J\sim \sqrt{\epsilon_E(M_{\rm P}^2 +\xi\phi^2)}\times {\cal H}/E_c$, which
is much below the Planck scale, and hence disastrous.
Similar problems occur when the scalar-graviton-graviton
vertices before partial integrations are considered.
Therefore, for a more complete understanding of naturalness
it is of crucial importance to understand the role of
the boundary terms (on equal time hypersurfaces).

\section{Discussion}
\label{sec:Discussion}

 We have used the frame and gauge independent formalism
for scalar and tensor cosmological perturbations of Ref.~\cite{Prokopec:2013zya}
to show that the physical cutoff for 2-to-2 tree level scatterings in Higgs inflation
is above the Planck scale $M_{\rm P}=1/\sqrt{8\pi G_N}$ throughout inflation.
More precisely, we found that in the Jordan frame, the physical cutoff scale
is $(\Lambda/a)_J\gtrsim\sqrt{M_{\rm P}^2 +\xi\phi^2}$, while in
the Einstein frame it is $(\Lambda/a)_E\gtrsim M_{\rm P}$, where $\xi$ is the nonminimal coupling
and $\phi(t)$ denotes the Higgs {\it vev}. The physical cutoff in the Jordan frame is illustrated
in figure~\ref{fig:cutoffJordan}. The difference between the two frames
is immaterial in that it can be fully attributed to the frame dependence of
the (physical) cutoff, see Eq.~(\ref{scale factor in two frames}).
Our results are incomplete, in that we have not discussed the relevance of:
\begin{itemize}
\item[$\bullet$] quartic vertices and loops,
\item[$\bullet$] vertices containing gauge and fermionic fields,
\item[$\bullet$] boundary terms on equal time hypersurfaces (that result from partial integrations),
\end{itemize}
for the question of naturalness in Higgs inflation.
We do however believe that the principal conclusion reached in this paper will not change
when these contributions are fully accounted for.


\section*{Acknowledgements}

 We thank Damien George, Sander Mooij and Marieke Postma
for useful discussions. This work was in part supported by Nikhef, by the D-ITP consortium, a
program of the NWO that is funded by the Dutch Ministry of Education,
Culture and Science (OCW) and by the Institute for Theoretical Physics of Utrecht University.

%
%

\bibliographystyle{apsrev}


\end{document}